\documentclass[3p,times,procedia]{elsarticle}
\usepackage{nupha_ecrc}

\volume{00}

\firstpage{1}

\journalname{Nuclear Physics A}

\runauth{Jiangyong Jia on behalf of the ATLAS\fnref{col1} Collaboration}

\jid{nupha}

\jnltitlelogo{Nuclear Physics A}

\usepackage[colorlinks,breaklinks,linkcolor=blue,citecolor=blue]{hyperref}
\usepackage{amssymb}
\usepackage{wrapfig}
\newcommand{\pp}{\mbox{$pp$}}
\newcommand{\pPb}{\mbox{$p$+Pb}}
\newcommand{\sqrtsnn}{\sqrt{s_{\mathrm{NN}}}}

\newcommand{\pT}{\mbox{$p_{\rm{T}}$}}

\newcommand{\lr}[1]{\left\langle #1\right\rangle}

\newcommand{\Deta}{\mbox{$\Delta \eta$}}

\newcommand{\npart}{N_{\mathrm {part}}}

\newcommand{\nchrec}{N_{\mathrm{ch}}^{\mathrm{rec}}}
\newcommand{\nch}{N_{\mathrm{ch}}}

 \usepackage{lineno}

\usepackage[figuresright]{rotating}

\begin{document}

\begin{frontmatter}




\title{Forward-backward multiplicity correlations in $\pp$, $\pPb$ and Pb+Pb collisions with the ATLAS detector}

\author{Jiangyong Jia on behalf of the ATLAS\fnref{col1} Collaboration}
\fntext[col1] {A list of members of the ATLAS Collaboration and acknowledgements can be found at the end of this issue.}
\address{Chemistry Department, Stony Brook University, NY 11794 and Physics Department, Brookhaven National Laboratory, NY 11796, USA}

\begin{abstract}
Two-particle pseudorapidity correlations are measured in $\sqrt{s_{\rm{NN}}}$~=~2.76~TeV Pb+Pb, $\sqrt{s_{\rm{NN}}}$~=~5.02~TeV $\pPb$ and $\sqrt{s}$~=~13 TeV $\pp$ collisions~\cite{ATLAS}. Correlation function is measured using charged particles in the pseudorapidity range $|\eta|<2.4$ with transverse momentum $\pT>0.2$ GeV, and it is measured as a function of event multiplicity, defined by number of charged particles with $|\eta|<2.5$ and $\pT>0.4$ GeV. The correlation function is decomposed into a short-range component (SRC) and a long-range component (LRC). The SRC differs significantly between the opposite-charge pairs and same-charge pairs, and between the three collision systems at similar multiplicity. The LRC is described approximately by $1+\lr{a_1^2}\eta_1\eta_2$ in all collision systems over the full multiplicity range. The values of $\lr{a_1^2}$ are consistent between the opposite-charge and same-charge pairs, and are similar for the three collision systems at similar multiplicity. The values of $\lr{a_1^2}$ and the magnitude of the SRC both follow a power-law dependence on the event multiplicity.
\end{abstract}

\begin{keyword}
heavy ion\sep pseudorapidity correlation \sep short-range correlation \sep long-range correlation

\end{keyword}

\end{frontmatter}


\section{Introduction}\label{sec:1}
Two-particle correlations in azimuthal angle and pseudorapidity ($\eta$) is an valuable tool to study the space-time dynamics of the heavy ion collisions. One important recent result is the observation of the long-range ridge correlation in $\pp$ and $\pPb$ collisions, with a magnitude comparable to those observed in Pb+Pb collisions at similar multiplicity~\cite{ridge}. There are presently significant debates on the origin of the apparent collectivity in small collision systems, whether this is a final state effect signifying a QGP-like matter or this is a initial state effect associated with strong QCD field. To clarify the situation, it is crucial to understand the nature of the sources of particle production that seed these long-range collective ridges: What are these sources made of and how many? What are their sizes and distribution in the transverse direction?

Forward-backward (FB) multiplicity correlation provides a handle on these questions. Due to quantum fluctuations of the nuclear wavefunction, the number of colliding objects, participating nucleons or in general colliding partons, in the target and those in the projectile are not the same on the event-by-event (EbyE) basis~\cite{Jia:2014ysa}. This asymmetry leads to correlation of particles with large pseudorapidity separation (long-range correlations or LRC). The colliding objects also serve as the sources for particle production which drive the transverse flow dynamics: the eccentricity of a collision depends on the distribution of the sources in the transverse plane, and it usually decreases with the number of sources $n$ as $\epsilon_m\propto 1/\sqrt{n}$. Therefore measuring the LRC also provides important information on the initial condition for the transverse collective dynamics in $\pp$, $\pPb$ and Pb+Pb collisions~\cite{schenke}.

Many previous studies in this direction are based on FB correlations of particle multiplicity in two $\eta$ ranges symmetric around the centre-of-mass of the collision system. This study is based on measuring a simple 2-D correlation function in $\eta$~\cite{Berger:1974vn,Jia:2014ysa}:
\begin{eqnarray}
\label{eq:1}
C(\eta_1,\eta_2) = \frac{\left\langle N(\eta_1) N(\eta_2)\right\rangle}{\left\langle N(\eta_1)\right\rangle\left\langle N(\eta_2)\right\rangle}\equiv \left\langle R(\eta_1)R(\eta_2)\right\rangle\;,\;\;\; R(\eta)\equiv \frac{ N(\eta)}{\left\langle N(\eta)\right\rangle}\;,
\end{eqnarray}
where $N(\eta)\equiv~\mathrm{d} N/\rm{d}\eta$ is the multiplicity density distribution in a single event and $\lr{N(\eta)}$ is the average distribution for a given event-multiplicity class. The correlation function is directly related to a single-particle quantity $R(\eta)$, which characterizes the fluctuation of multiplicity in a single event relative to the average shape of the event class. The $C(\eta_1,\eta_2)$ is constructed using the usual event-mixing method, and a renormalization procedure is used to remove the residual centrality dependence in $\left\langle N(\eta)\right\rangle$, with the resulting correlation function denoted by $C_{\rm N }(\eta_1,\eta_2)$~\cite{Jia:2014ysa}.

\vspace*{-0.4cm}\section{Results}\label{sec:2}
\begin{figure}[!h]
\centering
\includegraphics[width=0.8\linewidth]{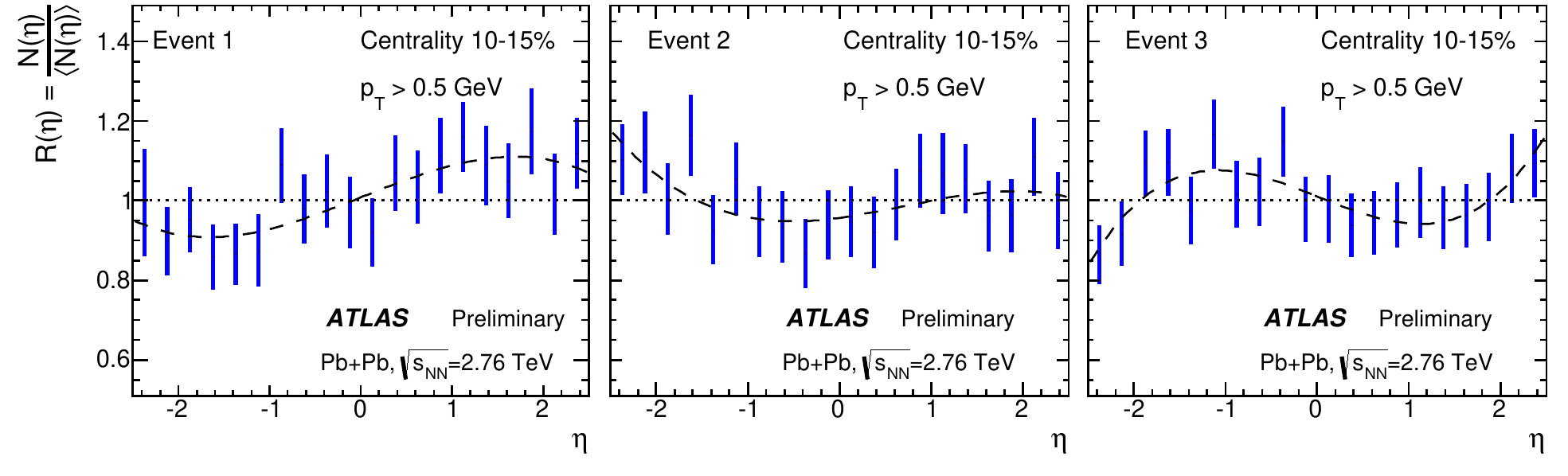}\vspace*{-0.3cm}
\caption{\label{fig:1} The multiplicity distributions of three typical events in 15–-20\% centrality interval. They are divided by the average distribution of all events in the same centrality interval. The dashed lines indicate fits to a third-order polynorminal function. Taken from Ref.~\cite{ATLAS}.}
\end{figure}

Figure~\ref{fig:1} shows the $R(\eta)$ in three typical Pb+Pb events in the 10-15\% centrality interval~\cite{ATLAS}. The apparent non-uniform structures reflect both the statistical fluctuations and dynamical fluctuations of interest. The advantage of the correlation function is that the statistical fluctuations natrually drop out after the averaging (eq.~\ref{eq:1}) and only the dynamical fluctuations remain.

\begin{wrapfigure}{r}{0.45\linewidth}
\centering
\vspace*{-0.4cm}
\includegraphics[width=1\linewidth]{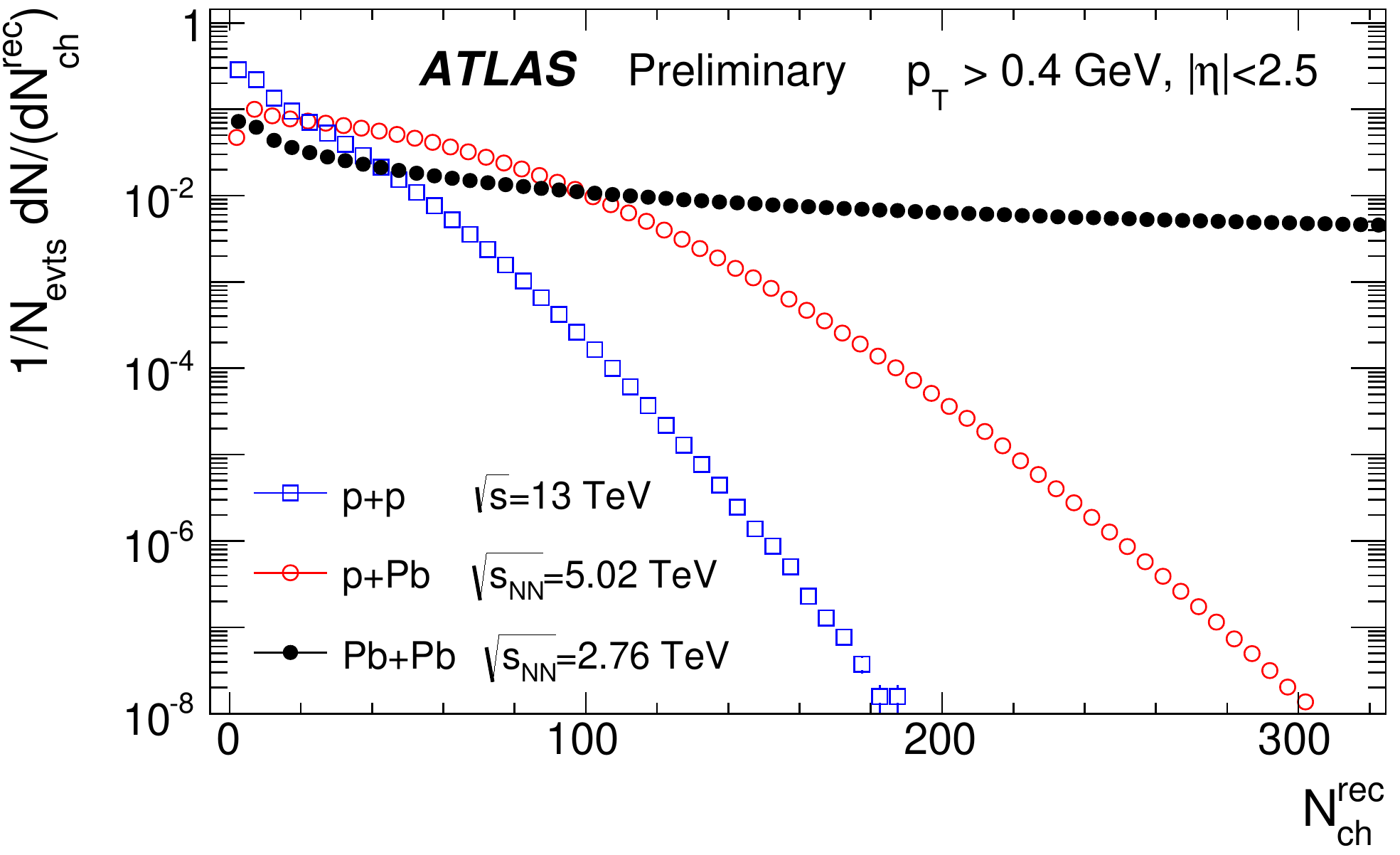}\vspace*{-0.3cm}
\caption{\label{fig:2}  The distributions of number of reconstructed tracks with $\pT > 0.4$ GeV and $|\eta|<2.5$ in the three collision systems. Taken from Ref.~\cite{ATLAS}.}
\vspace*{-0.2cm}
\end{wrapfigure}
The $C_{\rm N}(\eta_1,\eta_2)$ is measured in $\sqrt{s_{NN}}$=2.76 TeV Pb+Pb, $\sqrtsnn$=5.02 TeV $\pPb$ and $\sqrt{s}$=13 TeV $\pp$ collisions, using the ATLAS detector~\cite{ATLAS}. Events are classified according to the total number of reconstructed charged particles, $\nchrec$, with $|\eta|<2.5$ and transverse momentum $\pT>0.4$ GeV. The magnitudes of the FB fluctuations are compared for the three systems at similar event multiplicity.  As shown by Fig.~\ref{fig:2}, the range of $\nchrec$ distribution is much broader in Pb+Pb collisions than that in $\pPb$ collisions which in turn is much broader than that in the $\pp$ collisions. Therefore, the probability for events with large $\nchrec$ in $\pp$ collisions is much smaller than that in Pb+Pb collisions. One interesting question is whether the long-range multiplicity correlation is controlled by $\nchrec$ similar to the ridge phenomena or it also depends on other quantities. 

\begin{figure}[!h]
\centering
\includegraphics[width=0.8\linewidth]{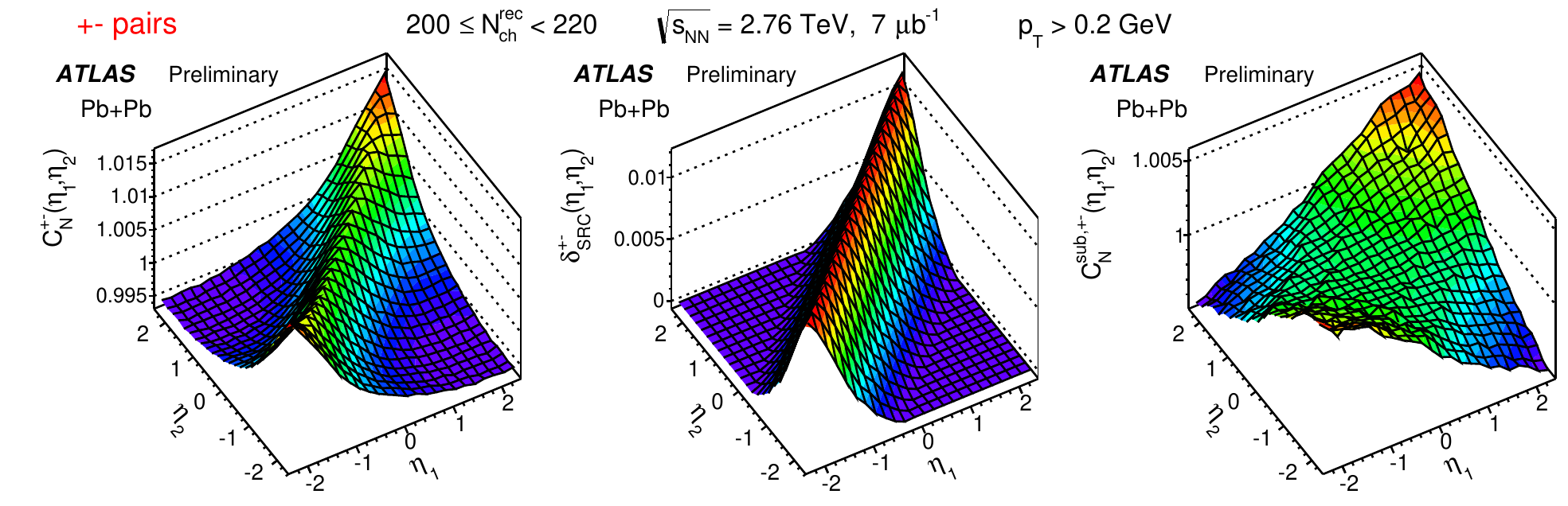}\\\vspace*{0.cm}
\includegraphics[width=0.8\linewidth]{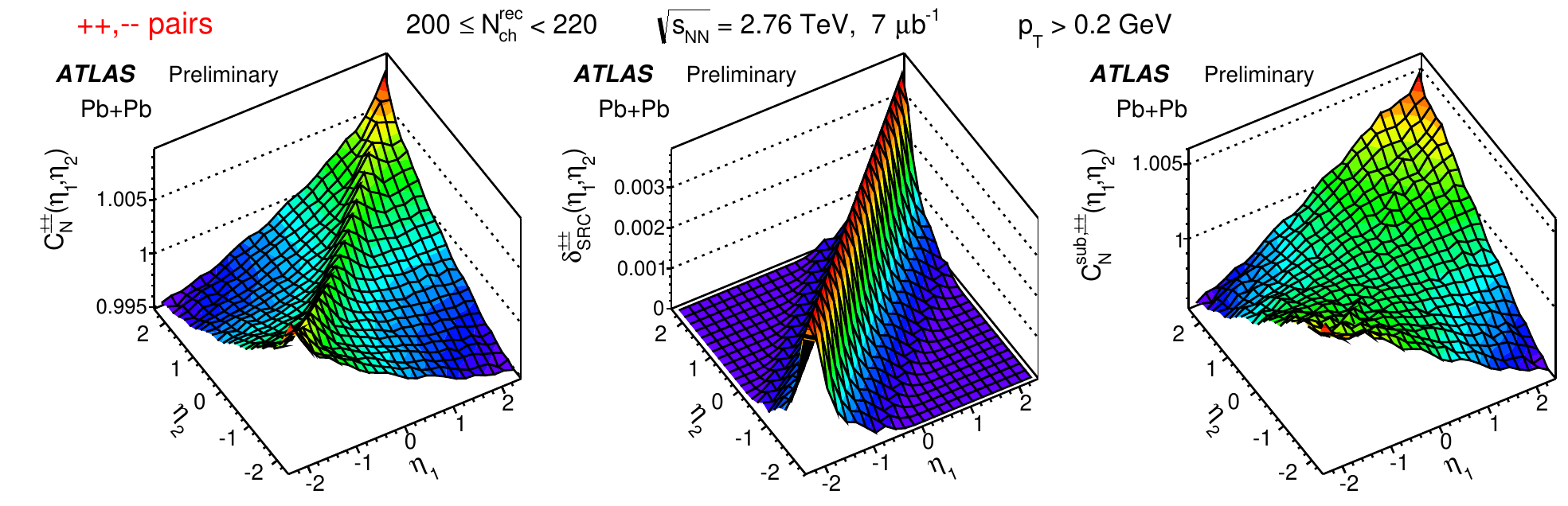}
\caption{\label{fig:3} The Pb+Pb correlation function $C_{N}(\eta_1,\eta_2)$ (left column), its short-range component (middle column) and long-range component (right column) in $200\leq \nchrec < 220$ multiplicity range for opposite-charge pairs (top) and same-charge pairs (bottom). Taken from Ref.~\cite{ATLAS}.}\vspace*{-0.4cm}
\end{figure}

Figure~\ref{fig:3} shows charge dependent Pb+Pb correlation functions in $200\leq \nchrec < 220$ multiplicity range. The correlation functions show a broad peak along $\Deta\sim0$ associated with short-range correlations (SRC), and a depletion at large $|\Deta|$ associated with LRC. The SRC reflects correlations within the same source, while the LRC reflects FB-asymmetry of the number of sources. The magnitude of SRC differs by more than factor of three between same-charge and opposite-charge pairs, while the LRC are nearly identical between the two charge combinations. Based on this, a data-driven method was developed to separate the SRC from LRC. The resulting SRC denoted by $\delta_{\rm{SRC}}(\eta_1,\eta_2)$ and LRC denoted by $C_{\rm{N}}^{\mathrm{sub}}(\eta_1,\eta_2)$ are shown in the middle and right columns, respectively. The SRC extends nearly $\pm 1-2$ in $\Deta$ independent of $\eta_+\equiv\eta_1+\eta_2$, and its magnitude is quantified by:
\vspace*{-0.4cm}\begin{eqnarray}\label{eq:2a}
\Delta_{\mathrm{SRC}} = \frac{\int \delta_{\rm {SRC}}(\eta_1,\eta_2) d\eta_1d\eta_2}{4Y^2}\;, |\eta|<Y=2.4\;.
\end{eqnarray}\vspace*{-0.1cm}
The saddle-like shape of the LRC is found to be well approximated by:\vspace*{-0.1cm}
\begin{eqnarray}\label{eq:2b}
C_{\rm N}^{\rm {sub}}\approx 1+\lr{a_1^2}\eta_1\eta_2,
\end{eqnarray}
suggesting that the dynamical fluctuation of $R(\eta)$ is linear in $\eta$. 

\begin{figure}[!h]
\centering
\includegraphics[width=0.8\linewidth]{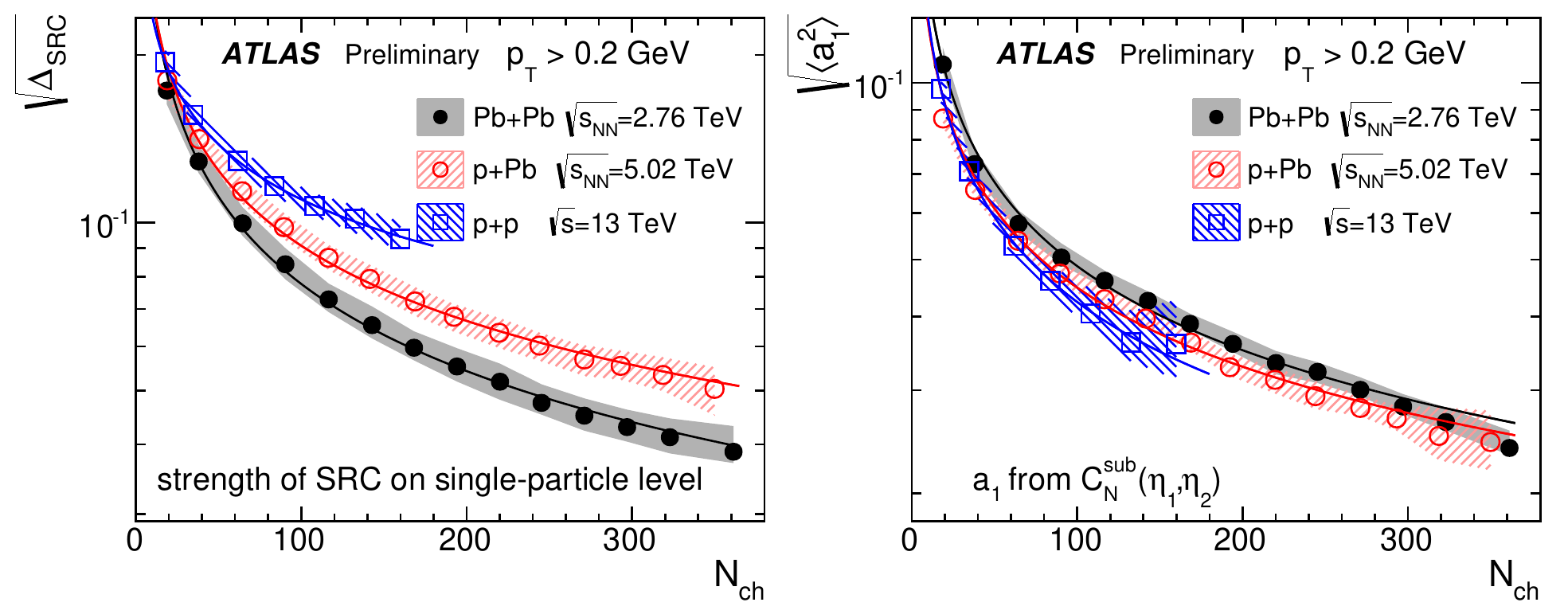}\vspace*{-0.4cm}
\caption{\label{fig:4}  The estimated magnitude of the SRC $\sqrt{\Delta_{\rm {SRC}}}$ (left panel) and LRC and $\sqrt{\lr{a_1^2}}$ (right panel) as a function of $\nch$ for all-charge pairs in Pb+Pb (solid circles), $\pPb$ (open circles) and $\pp$ (open squares) collisions. Taken from Ref.~\cite{ATLAS}.}\vspace*{-0.4cm}
\end{figure}

Figure~\ref{fig:4} compares the strength of SRC in terms of $\sqrt{\Delta_{\mathrm{SRC}}}$ and LRC in terms of $\sqrt{\lr{a_1^2}}$ between the three collision systems as a function of $\nch$ (the efficiency corrected $\nchrec$). All distributions follow a simple power law dependence on $\nch$. However, the magnitude of SRC is much stronger in $\pp$ collisions than in Pb+Pb collisions; in contrast the LRC signal is found to be similar between the three collision systems. This is a nontrivial result as the distribution of $\nch$ as well as the sizes of the overlap region are very different between the three collision systems.

The strength of the SRC and LRC is related to the number of sources $n$ contributing to the final multiplicity $\nch$, which is the sum of the number of sources from the projectile and target nucleon or nucleus, $n=n_{\rm f}+n_{\rm b}$. The LRC is expected to be related to the asymmetry between $n_{\rm f}$ and $n_{\rm b}$ :$A_n = \frac{n_{\rm f}-n_{\rm b}}{n_{\rm f}+n_{\rm b}},\;\lr{a_1^2} \propto \lr{A_n^2}$. The sources may be consists of participating nucleons $\npart$, sub-nucleonic degrees of freedom such as the fragmentation of scattered partons, or resonance decays. In an independent cluster model scenario~\cite{Berger:1974vn}, each source emits the same number of pairs and the number of sources follows a Poisson distribution. In this picture, the strength of SRC and LRC should scale approximately as the inverse of the number of sources. Therefore, assuming $n\propto\nch$, the $\sqrt{\Delta_{\rm {SRC}}}$ and $\sqrt{\lr{a_1^2}}$ in Fig.~\ref{fig:4} are expected to follow a simple power-law function in $\nch$:
\begin{eqnarray}\label{eq:e4}
\sqrt{\Delta_{\rm {SRC}}} \sim \sqrt{\lr{a_1^2}} \sim \frac{1}{n^{\alpha}}\sim \frac{1}{\nch^{\alpha}}\;, \alpha\sim0.5
\end{eqnarray}
A power index that is less than 0.5, $\alpha<0.5$, would suggest that $n$ grows slower than $\nchrec$, and vice versa.

To test this idea, the data in Fig.~\ref{fig:4} are fit to a power-law function: $c/\nch^{\alpha}$. The extracted power index values are summarized in Table~\ref{tab:fit}. The values of $\alpha$ for SRC are found to be smaller for smaller collision systems, they are close to 0.5 in Pb+Pb collisions and are significantly smaller than 0.5 in $\pp$ collisions. In contrast, the values of $\alpha$ for $\sqrt{\lr{a_1^2}}$ agree within uncertainties between the three colliding systems and are slightly below 0.5. 
\begin{table}[!h]
\centering
\begin{tabular}{c|c|c|c}\hline
                                        & Pb+Pb            & $\pPb$         & $\pp$            \tabularnewline\hline
$\alpha$ for $\sqrt{\Delta_{\rm {SRC}}}$  & $0.502\pm0.022$ & $0.451\pm0.020$ & $0.342\pm0.030$ \tabularnewline\hline
$\alpha$ for $\sqrt{\lr{a_1^2}}$         & $0.467\pm0.011$ & $0.448\pm0.019$ & $0.489\pm0.032$ \tabularnewline\hline
\end{tabular}\vspace*{-0.2cm}
\caption{\label{tab:fit} The power index and associated total uncertainty from a power-law fit of the $\nch$ dependence of $\sqrt{\Delta_{\rm {SRC}}}$ and $\sqrt{\lr{a_1^2}}$. Taken from Ref.~\cite{ATLAS}.}
\end{table}

\vspace*{-0.9cm}\section{Summary}\label{sec:3}
Two-particle pseudorapidity correlations are measured in $\sqrt{s_{NN}}$ = 2.76 TeV Pb+Pb, $\sqrt{s_{NN}}$ = 5.02 TeV $\pPb$ and $\sqrt{s}$ = 13 TeV $\pp$ collisions. The correlation function $C_{\rm N}(\eta_1,\eta_2)$ is measured using charged particles in the pseudorapidity range $|\eta|<2.4$ with transverse momentum $\pT>0.2$ GeV, and is decomposed into a short-range component (SRC) and a long-range component (LRC). The SRC is centered around $\Deta\sim0$ with a width of 1-2 units, while LRC has an approximate functional form $1+\lr{a_1^2}\eta_1\eta_2$. The magnitudes of the SRC and LRC are compared between the three collision systems as a function of $\nch$. Large differences are observed for the SRC, but the strength of the LRC agrees within $\pm10$--20\% at the same $\nch$. The $\nch$ dependences of both SRC and $\lr{a_1^2}$ follow an approximate power-law shape. The power index for $\lr{a_1^2}$ is approximately the same between the three collision systems. In contrast, the power-law index for the SRC is smaller for smaller collision systems. 

This research is supported by NSF under grant number PHY-1305037 and by DOE through BNL under contract number DE-SC0012704.
\vspace*{-0.2cm}







\end{document}